\documentclass[twocolumn,amsmath,amssymb,aps,prb,nofootinbib,floatfix]{revtex4-2}
\usepackage[dvipsnames]{xcolor}
\definecolor{p-channel}{HTML}{78BCD8}
\definecolor{c-channel}{HTML}{D36146}
\definecolor{d-channel}{HTML}{66458C}
\usepackage[
  colorlinks,
  linkcolor=c-channel!60!black,
  citecolor=p-channel!60!black,
  urlcolor=d-channel!80!black,
]{hyperref}
\usepackage{graphicx}
\usepackage{bm}
\usepackage{tikz}
\usepackage{mathtools}
\usepackage{braket}
\usepackage[capitalize]{cleveref}
\usepackage{comment}

\usepackage{makerobust}
\newcommand{\makeauthor}[2]{\newcommand{#1}[1]{{%
  \protect%
  \color{#2}{%
    \bfseries%
    \begingroup\escapechar=-1\edef\x{\endgroup\string#1}\x:%
  }\itshape{} ##1}}%
  \MakeRobustCommand#1}
\makeauthor{\lk}{RedViolet}
\makeauthor{\md}{YellowOrange}
\makeauthor{\ab}{BlueGreen}
\newcommand{\bv}[1]{\boldsymbol{#1}}
\newcommand{\dd}{\mathrm{d}}
\let\eqref\cref

\let\oldcite\cite
\renewcommand{\cite}[1]{\if\relax\detokenize{#1}\relax\textbf{\color{red}[?]}\else\oldcite{#1}\fi}

\begin{document}

\title{Electron-Phonon Functional Renormalization Group of Fermi Liquid Instabilities}

\author{C.~Alexander Baum}
\email{christian.baum@uni-wuerzburg.de}
\affiliation{Institut für Theoretische Physik und Astrophysik and Würzburg-Dresden Cluster of Excellence ctd.qmat, Universität Würzburg, 97074 Würzburg, Germany}
\author{Matteo Dürrnagel}
\affiliation{Institut für Theoretische Physik und Astrophysik and Würzburg-Dresden Cluster of Excellence ctd.qmat, Universität Würzburg, 97074 Würzburg, Germany}
\author{Ronny Thomale}
\affiliation{Institut für Theoretische Physik und Astrophysik and Würzburg-Dresden Cluster of Excellence ctd.qmat, Universität Würzburg, 97074 Würzburg, Germany}
\author{Lennart Klebl}
\email{lennart.klebl@uni-wuerzburg.de}
\affiliation{Institut für Theoretische Physik und Astrophysik and Würzburg-Dresden Cluster of Excellence ctd.qmat, Universität Würzburg, 97074 Würzburg, Germany}

\date{\today}

\begin{abstract}
We formulate a functional renormalization group (FRG) ansatz for correlated electron models that incorporates electronic interactions as well as electron-phonon coupling (EPC) stemming from dispersive phonon bands. Particularizing to the RG flow of the electron-electron interaction vertex, we treat phonon- and electron-mediated Fermi liquid instabilities on equal footing as we analyze tentative electronic order parameters related to charge, spin, nematicity, and superconducting pairing. We illustrate the approach at the example of Peierls-type transitions we find for the Hubbard model on the square lattice coupled to acoustic phonon bands.  Our method allows to incorporate full electronic and phononic \emph{ab initio} input, and thus lends itself to the analysis of electronic order from intertwined electronic interactions and EPC at a microscopically most substantiated level.      
\end{abstract}

\maketitle

\section{Introduction}
Facing the overwhelming complexity and intricacy of correlated electron systems at scales small against the atomic, crystalline, and Fermi energy scales, approximative organization principles of the relevant remainder degrees of freedom are typically indispensable, and hence have entered the elementary theory repertoire of condensed matter.  
To address the interplay of electrons and phonons, for instance, a milestone effective modelling is given by the Born-Oppenheimer approximation (BOA)~\cite{Born1927}, where the ionic and electronic degrees of freedom are separated so as to treat their dynamics independently. It is based on the assumption that the electronic fluctuations have a much shorter characteristic timescale compared to the slow distortions of the crystal lattice, and thus adiabatically follow any change in lattice potential. While the crosstalk between the electronic sector and quantized lattice vibrations has proven essential to describe lattice deformations akin to the Kohn anomaly~\cite{Kohn1959i} and the Jahn-Teller effect~\cite{Jahn1937s}, electrons are treated as only acting as non-interacting screening background, while the charge ordering is primarily governed by the momentum structure of the associated electron phonon coupling (EPC)~\cite{Johannes2008}. By contrast, time reversal symmetry breaking instabilities such as magnetic orders are predominantly addressed in the electronic sector, as they are usually not accompanied by a concomitant lattice reconstruction.

This conventional wisdom is challenged by an increasing class of materials including kagome metals~\cite{Jiang2021,Zhao2021,Mielke2022,DiSante2026}. There, electronic correlations are believed to play a crucial role in the observed exotic charge orders~\cite{Kiesel2013u, Neupert2022, Huai2025, Wang2021}, where their interplay with lattice degrees of freedom can lead to the emergence of orbital magnetic moments and chiral phonons~\cite{zhan2025loop, Chen2025}.
Conversely, the orbital magnetic moment of chiral phonon fluctuations can significantly alter the magnetic ordering expected from a purely electronic mechanism~\cite{Juraschek2019,Luo2023,Tonacatl2025}.
While the partially hierarchical treatment of both sectors inspired by BOA might still be justified in the aforementioned instances, electronic scenarios found in CsCr\textsubscript3Sb\textsubscript5~\cite{liu2024superconductivity, Liu2026} or the trilayer nickelate La\textsubscript4Ni\textsubscript3O\textsubscript{10}~\cite{Zhang2020, Khasanov2025}
with inherently coupled charge and magnetic orders necessitate a unified, i.e., non-separate treatment of electronic and lattice degrees of freedoms to faithfully decipher the microscopic nature of electronic order at low energies.

Previous attempts to investigate this interplay in simplified toy models include studies employing Eliashberg theory~\cite{Eliashberg1960, Marsiglio1990,Marsiglio2020},
quantum Monte Carlo~\cite{Scalettar1989,Noack1991, Sengupta2003,Costa2023,Feng2022, Assaad2007,Esterlis2018, Götz2022, Götz2024, cohenstead2026smoqyelphqmcjl}
dynamical mean field theory~\cite{Werner2007},
and various renormalization group approaches~\cite{Jeckelmann1998,Banerjee2023, Wang2015,Yang2022,Yirga2023}, but were limited to low dimensional systems with trivial local Hilbert space and few phonon modes such as generalized Hubbard-Holstein~\cite{Holstein1959} or Su-Schrieffer-Heeger models~\cite{Su1979}. As a tendency, these limitations have hitherto evaded the formulation of microsopic ans\"atze for quantum materials that would allow to treat emanating electronic orders on equal footing in the presence of EPC and electronic interactions.

In this work, we present a unified treatment based on the functional renormalization group (FRG)~\cite{Metzner2012, Platt2013} which does allow to consider electronic and phononic degrees of freedom on equal footing, and, in terms of numerical performance, is computationally equivalent to established FRG formulations for all-electronic models. It allows to include (almost) arbitrarily many phonon bands and only mildly constrains the number of addressable electronic bands to the order of ten, and may hence be readily applied to \emph{ab initio} derived models.
This is achieved by integrating out the phonons to obtain an all-electronic theory with retarded interactions, such that the cooperation and competition between electronic correlations and electron-phonon coupling can be investigated by the renormalization of a single object: the electronic effective action.
This gives way to a consistent picture for lattice instabilities in line with earlier \textit{ab initio} works, that treat this problem from a phononic perspective~\cite{Berges2023p, Schobert2024a}. To only name one remarkable immediate consequence out of this methodological progress, it for instance enables us to describe the competition between conventional and unconventional superconducting order within one single formalism. Furthermore, even in the absence of significant electronic interactions, our procedure includes vertex corrections to the renormalization of the electron-phonon coupling, and therefore goes beyond the standard phonon softening mechanism based on the (bare) electronic charge susceptibility. Due to its straightforward scalability to higher dimensions and existing interfaces to \textit{ab initio} codes~\cite{10.21468/SciPostPhysCodeb.26, 10.21468/SciPostPhysCodeb.26-r0.5}, our study sets the stage for material studies of coupled electron phonon systems with realistic first principle Hamiltonians.

This Article is organized as follows. In order to illustrate our ansatz we particularize to the square lattice Hubbard model with acoustic phonons, and analyze the rich interplay between electronic and phononic interactions captured by our method.
First, we introduce the Hubbard model with acoustic phonons in \cref{sec:model} and explicitly derive the phonon modes with their corresponding electron-phonon coupling.
In \cref{sec:method} we sketch how to integrate out the phonons and include the resulting retarded interaction into the fermionic FRG within standard approximations.
We further show that a subset of the diagrams resummed in FRG reproduces the leading order contribution to phonon softening, and discuss how to connect the FRG approach to \textit{ab initio} methods such as density functional perturbation theory.
In \cref{sec:results}, we then apply this formalism to the acoustic phonons in the square lattice Hubbard model.
We show how softening of phonon modes and charge-bond order from the electronic picture condify the same transition from two different perspectives. 
We continue by combining electron-phonon coupling with Coulomb repulsion (Hubbard-$U$) through our FRG to show how the two effects intertwine to form a complex phase diagram, including both $s$ and $d$-wave superconductivity.
In \cref{sec:conclusion} we conclude that our electron-phonon functional renormalization group ansatz has significant potential in refining our understanding of electronic order and its interplay with electronic interaction and electron-phonon coupling at a microscopically substantiated level. 

\begin{figure*}[!t]
    \centering
    \includegraphics{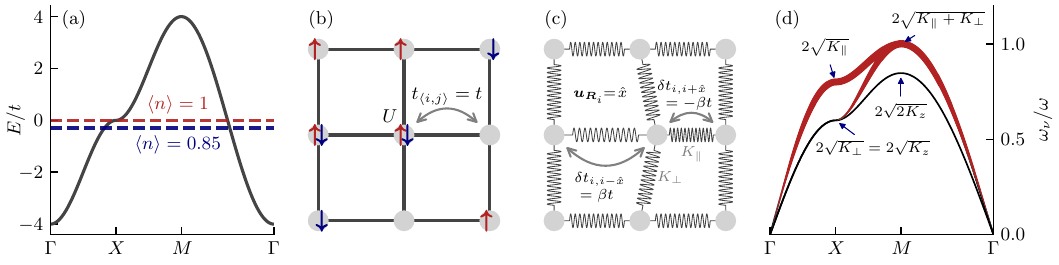}
    \caption{The Hubbard model on the square lattice~(b) with its band structure~(a). We indicate the Fermi level at half filling $\langle n\rangle = 1$ as dashed red line, and the slightly doped case $\langle n \rangle = 0.85$ as dashed blue line. Panel~(c) presents a schematic for the phonon spring model and the electron phonon coupling induced by a local perturbation at site $\bv R_i$. The Phonon band structure is depicted in~(d), showing how the spring constants $K_\parallel$, $K_\perp$, $K_z$ influence the phonon dispersion.
    The modes with longitudinal component have non-zero electron-phonon coupling [red, line thickness corresponding to its strength $s_{\bv q}^{(\nu)} = \sum_{j}|g_{\bv q, 0j}^{(\nu)}|^2$, cf.~\cref{eq:momentum_space_EPC}], while the shear modes have no coupling (black, red line thickness zero).
    }
    \label{fig:model}
\end{figure*}

\section{Model} \label{sec:model}

The full Hamiltonian of a combined electron-phonon system can be split into three parts
\begin{equation}
    H = H_\mathrm{el} + H_\mathrm{ph} + H_\mathrm{elph} \,.
    \label{eq:Hamiltonian}
\end{equation}
Here, $H_\mathrm{el}$ denotes the electronic Hamiltonian, $H_\mathrm{ph}$ the phononic Hamiltonian, and $H_\mathrm{elph}$ the electron-phonon coupling Hamiltonian. For the square lattice tight-binding model with on-site Hubbard interaction $U$, $H_\mathrm{el}$ reads
\begin{equation}
    H_\mathrm{el}= t \sum_{\sigma, \langle i,j\rangle}c^\dagger_{i\sigma} c_{j\sigma} + U\ \sum_{i} \ n_{i \uparrow} n_{i \downarrow} \,,
    \label{eq:H_e}
\end{equation}
where $c_{i\sigma}^{(\dagger)}$ annihilates (creates) an electron with spin $\sigma\in\{\uparrow,\downarrow\}$ at site $i$, and $n_{i\sigma} = c^\dagger_{i\sigma}c^{\vphantom{\dagger}}_{i\sigma}$. The sum extends over nearest neighbors $\langle i,j\rangle$ of the square lattice and results in the well known dispersion depicted in \cref{fig:model}(a) while the interaction term is a standard repulsive Hubbard interaction.

The phonons are assumed to be non-interacting, i.e., we work within the harmonic approximation. In the phonon band basis, $H_\mathrm{ph}$ can be generically written as
\begin{equation}
    H_\mathrm{ph} = \sum_{\nu, \bv{q}} \omega_{\nu,\bv{q}} \left(b^\dagger_{\nu, \bv{q}} b_{\nu, \bv{q}} + \frac12 \right) \,,
    \label{eq:H_ph}
\end{equation}
where $\nu$ is the phonon band index, $\bv{q}$ the bosonic momentum, $b_{\nu, \bv{q}}^{(\dagger)}$ the corresponding bosonic annihilation (creation) operator and $\omega_{\nu,\bv{q}}$ the phonon dispersion.

Lastly, the electron-phonon coupling in general reads
\begin{equation}
    H_\mathrm{elph} = \sum_{\sigma, i, j, \nu, \bv{q}} \frac{g^{(\nu)}_{\bv{q}, ij} }{\sqrt{2M\omega_{\nu,\bv q}}} \, e^{i \bv R_i \bv{q}} \big( b^\dagger_{\nu,-\bv q} + b_{\nu,\bv{q}} \big) c_{i\sigma}^\dagger c_{j\sigma}^{\vphantom{\dagger}} \,,
    \label{eq:H_elph}
\end{equation}
where $\bv R_i$ denotes the Bravais lattice vector corresponding to site $i$.
For the remainder of this work, we will study the square lattice Hubbard model with acoustic phonons visualized in \cref{fig:model}. In the following two subsections we will provide a detailed derivation of both the phonon dispersion $\omega_{\nu,\bv q}$ and the electron-phonon coupling $g^{(\nu)}_{\bv q, ij}$ of this system, which can easily be generalized to other minimal lattice models.

\subsection{Phonon modes} \label{sec:phonon_modes}

Within the harmonic approximation, the energy of lattice deformations is defined by interatomic force constants (IFC):
\begin{equation}
    C_{\bv R-\bv R',ij} = \frac{\partial^2 E}{\partial u_{\bv R,i}\partial u_{\bv R',j}}\bigg|_{u \equiv 0} \,,
    \label{eq:ifcs}
\end{equation}
that give the force acted on an atom at $\bv R$ in direction $i$ from an atom at $\bv R'$ perturbed in direction $j$.%
\footnote{For more complicated systems, the directional indices $i,j$ also label the sublattice inside the unit cell.}
The dynamical matrix is defined as the Fourier transform of the IFC corrected for the masses of the nuclei:
\begin{equation}
    D_{\bv{q},ij} = \frac{1}{\sqrt{M_iM_j}}\sum_{\bv R} C_{\bv R,ij} e^{i \bv{q} \bv R} \,,
    \label{eq:dynamic_matrix}
\end{equation}
where $M_i = M_j = M$ is the mass of each atom.
The phononic dispersion $\omega_{\nu,\bv{q}}$ and modes $\epsilon_{\nu,i}(\bv q)$ are then calculated via diagonalization in $ij$:
\begin{equation}
    D_{\bv{q},ij} \, \epsilon_{\nu,j}(\bv q) = \omega^2_{\nu,\bv{q}} \,\epsilon_{\nu,i}(\bv q) \,,
    \label{eq:ph_eigenproblem}
\end{equation}
from which the phonon operators can be defined through~\cite{Berges2020}
\begin{equation}
    u_{\nu,\bv{q}} = \frac{1}{\sqrt{N_{\bv R}}}\sum_{l} \, \epsilon^{*}_{\nu,l}(\bv q) \sum_{\bv R} u_{\bv R, l} \, e^{i\bv{q}\bv R} \,,
    \label{eq:ph_displacement_q}
\end{equation}
and
\begin{equation}
    u_{\nu,\bv q} = \frac{1}{\sqrt{2M\omega_{\nu \bv q}}} \big[b^\dagger_{\nu,-\bv q} + b_{\nu,\bv q}^{\vphantom{\dagger}} \big] \,,
    \label{eq:ph_operator}
\end{equation}
which defines the bosonic operators found in \eqref{eq:H_ph}.

For the IFC on the square lattice we only consider nearest-neighbor springs [cf.~\cref{fig:model}(c)]. $C_4$ rotation symmetry implies
\begin{equation}
    C_{\hat x} = C_{-\hat x}, \quad C_{\hat y} = C_{-\hat y} = R_{\pi/2} C_{\hat x} R_{-\pi/2} \,,
    \label{eq:ifcs_square}
\end{equation}
where $C_{\hat x, ij} = \mathrm{diag}(K_\parallel, K_\perp, K_z)$ entails the spring constants for longitudinal, transversal and out-of-plane displacement w.r.t.~the spring axis and $R_\varphi$ is a $\varphi$-rotation around the $\hat z$-axis.
Since in equilibrium all atoms are at their unperturbed position, the acoustic sum rule requires the dynamical matrix to identically vanish at $\bv q=0$. This fixes the on-site component to $C_{0} = - 2(C_{\hat x} + C_{\hat y})$ and the resulting dynamical matrix,
\begin{equation}
    M D_{\bv{q}} = 2\left[\cos(q_x)-1\right] C_{\hat x} + 2\left[\cos(q_y)-1\right] C_{\hat y} \,,
    \label{eq:dynamical_matrix_square}
\end{equation}
has only acoustic modes [cf.~\cref{fig:model}(d)].
The eigendecomposition yields one out-of-plane mode and two in-plane modes
\begin{equation}
\begin{aligned}
    M\omega_{z}^2 &{}= -2 K_z \left[\cos(q_x) + \cos(q_y)-2\right] \,,\\
    M\omega_{x/y}^2 &{}= -2 K_\parallel \left[\cos(q_{x/y})-1\right] - 2 K_\perp \left[\cos(q_{y/x})-1\right] \,,
    \label{eq:ph_modes_square}
\end{aligned}
\end{equation}
with trivial eigenvectors $\epsilon_{\nu,i}(\bv q) = \hat \nu$ for $\nu\in\{x,y,z\}$.

\subsection{Electron-Phonon coupling} \label{sec:electron_phonon_coupling}
With $\omega_{\nu \bv q}$ obtained from \cref{eq:ph_modes_square}, we complete our derivation of the model Hamiltonian by determining the precise form of the electron phonon coupling (EPC) $g^{(\nu)}_{\bv q, ij}$ in \cref{eq:H_elph}.
The EPC can be derived from the perturbation of the electronic single-particle Hamiltonian given a local lattice displacements $u_{\bv R,l}$:
\begin{equation}
    g^{l}_{\bv{R}, ij}  =  \frac{\partial t_{ij}}{\partial u_{\bv{R},l}}\bigg|_{u \equiv 0} \,.
    \label{eq:elph_general}
\end{equation}
By starting from a tight-binding picture, we have implicitly assumed exponentially localized electronic orbitals.
The hoppings (which are their overlap integrals) inherit the exponential dependence on the inter-atomic distance, i.e.,
\begin{equation}
    t_{ij} \approx t_0 \exp\big[-\beta \big(|\bv R_i + \bv u_{\bv R_i} - \bv R_j - \bv u_{\bv R_j} |-a\big)\big] \,,
    \label{eq:tij_exp}
\end{equation}
where $\beta$ parametrizes the orbital spread.
To linear order in $\bv u$, the change in hopping is then
\begin{equation}
    t_{ij} - t_0 = -\beta t_0 \, \bv r_{ij} \cdot ( \bv u_{\bv{R}_i} - \bv u_{\bv{R}_j} ) + \mathcal O\big((\bv u_{\bv R_i} - \bv u_{\bv R_j})^2\big) \,,
    \label{eq:delta_tij}
\end{equation}
and $\bv r_{ij}$ is a normalized vector along the direction of the bond. 
\Cref{fig:model}(c) visualizes such a localized lattice perturbation at site $\bv R_i$ and the resulting electron phonon coupling.
Inside the approximation \eqref{eq:tij_exp}, the hopping $t_{ij}$ is only perturbed by displacement of either nuclei $\bv u_{\bv R_{i/j}}$ of the bond along the bond's direction, resulting in a naturally local representation of the EPC \eqref{eq:elph_general}.
Even for \emph{ab initio} scenarios, this relative localization persists, so that a description where both electronic and phononic degrees of freedom are in real space is the most compact way to define EPC~\cite{Giustino2007, Berges2020}.
To arrive at the EPC of a particular phonon mode in \cref{eq:H_elph} we weight all local lattice displacements with their contribution to the respective phononic mode:
\begin{equation}
    g^{(\nu)}_{\bv{q}, ij} = \sum_{l, \bv{R}} \epsilon_{\nu,l}(\bv q) \ g^l_{\bv{R}, ij}  \, e^{-i \bv R \bv{q}} \,.
    \label{eq:momentum_space_EPC}
\end{equation}

\begin{figure}
    \centering
    \vspace{-10pt}
    \begin{tikzpicture}[inner sep=0pt, outer sep=0pt]
    \node (a) at (0,0) {
        \includegraphics[width=1.0\columnwidth]{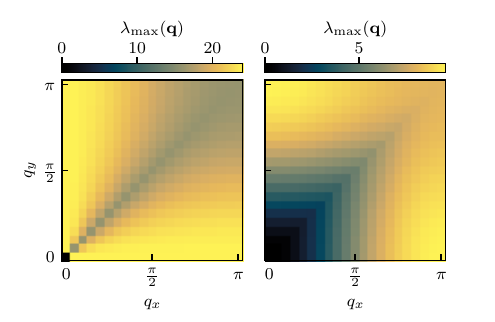}
    };
    \node[anchor=north east, xshift=55pt, yshift=15pt] at (a.south west) {\small(a)};
    \node[anchor=north west, xshift=15pt, yshift=15pt] at (a.south) {\small(b)};\textit{}
    \end{tikzpicture}
    \vspace{-18pt}
    \caption{Maximum eigenvalue of the phonon-mediated retarded interaction vertex \cref{eq:retarded_intn} as a function of
    $\bv q$ in the top right corner of the Brillouin zone. For $\omega_n = 0$~(a), there is a discontinuity at $\bv{q}=0$ as the $M \rightarrow \Gamma$ and $X \rightarrow \Gamma$ limits differ. This discontinuity does not occur for any nonzero transfer frequency, such as $\omega_n=1$~(b). The other parameters are given as $\lambda = 1$, $\omega = 1$ and $K_\perp = K_\parallel/2$ (cf.~\cref{sec:results} for definitions).
    }
    \label{fig:V_eff}
\end{figure}

\section{Methods} \label{sec:method}
\begin{figure*}[!t]
    \centering
    \begin{tikzpicture}
    \node (a) at (0,0) {
        \includegraphics[width=1.0\linewidth]{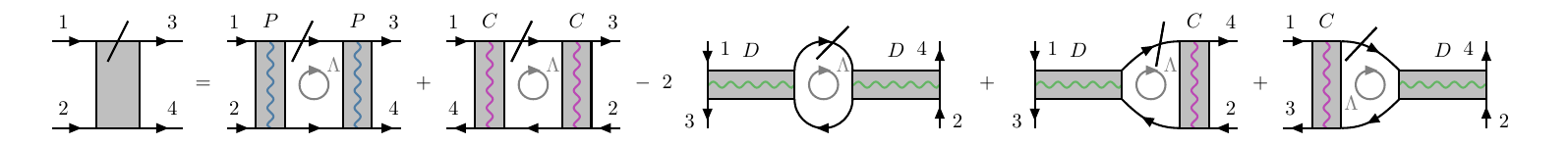}
    };
    \end{tikzpicture}
    \caption{The flow of the static vertex $V^{S,\Lambda}$ in channel decomposition.  The wiggly lines indicate the phonon propagator contained in the retarded vertex which contributes to the static part of the renormalized vertex with different frequencies for each channel [cf.~\crefrange{eq:static-vertex-mod-P}{eq:static-vertex-mod-D}].
    }
    \label{fig:flow_diagrams}
\end{figure*}
To treat the electron-electron interactions \cref{eq:H_e} and the electron-phonon coupling \cref{eq:H_elph} of the Hamiltonian \cref{eq:Hamiltonian} consistently, we employ the FRG. While it is in principle possible to treat coupled fermion-boson models via FRG directly~\cite{Berges2002} we prefer to work with a tried-and-tested all-electronic formalism~\cite{Metzner2012, Platt2013, Dupuis2021}.

In order to include electron-phonon couplings into the fermionic FRG we first integrate out the phonons to get a retarded interaction as done, e.g., in Refs.~\cite{Classen2014i, Wang2015, Yang2022}.
Further, we discuss the all-electronic Ansatz of the FRG with retarded interactions and compare to the self-energy corrections of the phonon modes in the case of charge order.
Importantly, the FRG in the direct particle-hole channel ($D$-channel) is equivalent to a random phase approximation (RPA) to the phonon self-energy, making the full FRG a natural extension of this approach that is unbiased in the fluctuation channels and can include electron-electron interactions.

\subsection{Phonon-induced retarded interaction}\label{ssec:retarded_interaction}
The induced action obtained from integrating out the phononic fields in \cref{eq:H_ph,eq:H_elph} reads (in imaginary time):
\begin{equation}
    S_\mathrm{ind} = -\frac{1}{2} \int_{\tau,\tau',\nu,\bv q} G^\mathrm{ph}_\nu(\bv{q}, \tau-\tau') \,\rho_\nu(\bv{q},\tau)\,\bar\rho_\nu(\bv{q},\tau') \,,
    \label{eq:induced_action}
\end{equation}
where
\begin{equation}
    G^\mathrm{ph}_\nu(\bv{q}, i\omega_n) = -\frac{2 \omega_{\nu,\bv{q}}}{\omega_n^2 + \omega_{\nu,\bv{q}}^2 - i \delta}
    \label{eq:ph_propagator}
\end{equation}
represents the phonon propagator in Matsubara frequency space with an infinitesimal regulator $\delta = 0^+$, and
\begin{equation}
    \rho_\nu(\bv{q},\tau') \equiv \sum_{i,j,\sigma} \frac{g^{(\nu)}_{\bv{q}, ij}}{\sqrt{2M\omega_{\nu,\bv q}}} \, e^{i \bv R_i\bv{q}} \, \bar c_{i\sigma}(\tau)c_{j\sigma}(\tau)
\end{equation}
is a generalized density bilinear of electronic fields.
We note that the regulator $\delta$ in \cref{eq:ph_propagator} takes care of the $\bv  q\to0$ limit and formally corresponds to a finite phonon lifetime. With the above definitions at hand, one may write the phonon-mediated electron-electron interaction as
\begin{equation}
    V_{ijkl}(\bv q,\omega_n) = \sum_{\nu} \frac{G^\mathrm{ph}_\nu(\bv{q}, i\omega_n)}{2M \omega_{\nu,\bv q}}\ \bar g^{(\nu)}_{\bv{q}, ij} g^{(\nu)}_{\bv{q}, kl} \ e^{i\bv{q} (\bv R_k-\bv R_i)} \,.
    \label{eq:retarded_intn}
\end{equation}
We see how the two electron bilinears $i,j$ and $k,l$ are coupled via the phononic momentum $\bv{q}$ in the effective vertex $V$.

Unlike for optical phonons, the $\bv{q}=0$ component in \cref{eq:retarded_intn} has a discontinuity at $\omega_n = 0$ for any acoustic phonon model as both the numerator and the denominator in the fraction $\bar g^{(\nu)}_{\bv{q}, ij} g^{(\nu)}_{\bv{q}, kl}/\omega_{\nu, \bv{q}}^2$ tend towards zero in a direction-dependent manner (see \cref{fig:V_eff}).
To alleviate this discontinuity we introduce a nonzero regulator $\delta = 10^{-7}$ for the numerical implementation, which automatically implies a vanishing contribution to the static interaction vertex at the zone center, i.e.,  $V_{ijkl}(\bv q = 0, 0) = 0$.
This provides a well defined starting point to apply the established all-electron FRG scheme to solve the combined electron-phonon system of \cref{eq:Hamiltonian}.

\subsection{Functional Renormalization}\label{ssec:FRG}
Following the usual procedure in electronic FRG~\cite{Lichtenstein2017} we decompose the combined two particle interaction of \cref{eq:H_e,eq:retarded_intn} into the bare part $V_0$ and the three diagrammatic channels:
\begin{multline}
    \label{eq:frg-channels}
    V^\Lambda (k_1, k_2; k_3, k_4) = V_0 + P^\Lambda(k_1 + k_2; k_1, k_3) \\
    {} + C^\Lambda(k_1 - k_4; k_1, k_3) + D^\Lambda(k_1 - k_3; k_1, k_4) \,,
\end{multline}
where the first momentum of $X^\Lambda(q; k, k')$ indicates the bosonic transfer momentum native to channel $X$ and the second two the remaining independent electronic momenta. Specifically, the particle-particle ($P$), crossed particle-hole ($C$) and direct particle-hole ($D$) channels correspond to the three different momentum transfers $s = k_1 + k_2$, $u = k_1 - k_4$, and $t = k_1 - k_3$, which we write as four-momentum Mandelstam variables, e.g., $t=(it_0,\bv t)$.
The phonon-induced retarded interaction of \cref{eq:retarded_intn} is native to the $D$-channel in the sense that it couples charge bilinears. Consequently, the transfer momentum (frequency) is equal to the phonon momentum (frequency) $\bv t = \bv{q}$ ($it_0 = i\omega_n$).

In the $SU(2)$ symmetric (spin rotationally invariant) case, the FRG flow equation of the two-particle vertex $V^\Lambda$ written in terms of the three diagrammatic channels then reads
\begin{align}
    \label{eq:frg-compact}
    \dot V^\Lambda &{}= \dot P^\Lambda + \dot C^\Lambda + \dot D^\Lambda \,, \\
    \label{eq:frg-compact-P}
    \dot P^\Lambda &{}= V^{\Lambda, P} \circ \dot \chi^{\Lambda, P} \circ V^{\Lambda, P} \,, \\
    \label{eq:frg-compact-C}
    \dot C^\Lambda &{}= V^{\Lambda, C} \circ \dot \chi^{\Lambda, H} \circ V^{\Lambda, C} \,, \\
    \label{eq:frg-compact-D}
    \dot D^\Lambda &{}= \begin{multlined}[t]
        -2\left( V^{\Lambda, D} - \frac12V^{\Lambda,C} \right) \circ \dot \chi^{\Lambda, H} \\ {}\circ \left( V^{\Lambda, D} - \frac12V^{\Lambda,C} \right) + \frac12 \dot C^\Lambda \,,
    \end{multlined}
\end{align}
where ``$\circ$''{} resembles a matrix product over all quantum numbers as expressed diagrammatically in \cref{fig:flow_diagrams}.
$V^{\Lambda,X}$ denotes the vertex projected to channel $X \in \{P,C,D\}$ and $\dot \chi^{\Lambda, P/H}$ is the particle-particle/particle-hole loop derivative:
\begin{equation}
\begin{aligned}
    \dot\chi^{\Lambda, P/H}(q,k) &{}= \dot G^\Lambda(k) G^\Lambda(\pm q\mp k) + \dot G^\Lambda(\pm q\mp k) G^\Lambda(k) \\
    &{}= \begin{multlined}[t]
    \frac12 \Big[ \delta(\Lambda - |ik_0|)
    \, G(k) G(\pm q\mp k) \\
    {} + \delta(\Lambda - |iq_0-ik_0|)\, G(\pm q\mp k) G(k) \Big] \,,
    \end{multlined}
\end{aligned}\label{eq:loopdev}
\end{equation}
with $G(k)$ the (bare) electronic Green's function. Note that \cref{eq:loopdev} uses a multiplicative sharp frequency cutoff as RG regulator, which dresses the electronic Green's function as follows:
$G^\Lambda(k) = \Theta(\Lambda - |ik_0|) G(k)$.
In \cref{eq:frg-compact-P,eq:frg-compact-C,eq:frg-compact-D}, the channel-specific Mandelstam variable remains diagonal, i.e., the corresponding matrix product becomes a scalar product. 
In order to reduce the numerical complexity of the flow \crefrange{eq:frg-compact}{eq:frg-compact-D}, we focus on the renormalization of the static vertex as outlined in the following.

\subsection{Flow equations with retarded interaction}\label{ssec:Flow_equation_retarded}
To separate the static from the retarded vertices along the RG flow, we employ the following decomposition~\cite{Wang2015, Yang2022, Zhan2025}:
\begin{multline}
    V^\Lambda(k_1,k_2;k_3,k_4) = V^{S,\Lambda}(\bv k_1, \bv k_2; \bv k_3, \bv k_4)\\
    + V^{R,\Lambda}(\{k_{i,0}\}, \{\bv k_{i}\}) \,.
\end{multline}
Here, $V^{S/R,\Lambda}$ corresponds to the static/retarded part and we insert the phonon-induced interaction \cref{eq:retarded_intn} for $V^{R,\Lambda}$:
\begin{equation}
    V^{R,\Lambda}(\{k_{i,0}\}, \{\bv k_{i}\}) = V_{ijkl}(\bv q,i\omega_n)\,.
\end{equation}
As we are primarily interested in a qualitative solution of the problem, i.e., the correct prediction of the leading symmetry breaking order and its dependence on the model parameters, we ignore the renormalization of the retarded interaction and consider it constant along the flow.\footnote{This renders the numerical complexity of the problem fully equivalent to static FRG, thus allowing for the application to realistic \emph{ab initio} models of quantum materials.}
We expect this approximation to yield qualitatively meaningful results, as the static part $V^{\mathrm{S}}$ is the decisive source for symmetry breaking in the all-electronic limit.
We further note that the (somewhat similar) Hubbard-Holstein model has been recently studied with FRG employing a full frequency-dependent flow~\cite{Al-Eryani2025}---with the results being fully in line with what has been found for the static approximation~\cite{Wang2015}.

In order to project the flow equation of the vertex onto the static part we set all external frequencies to zero and invoke frequency conservation.
Integrating over the remaining Matsubara frequency (we set $T=0$ such that the Matsubara axis is continuous) and making use of the functional form of the loop derivative \cref{eq:loopdev} leads to a vanishing transfer frequency (of the vertex) for the $D$-channel. 
Conversely, the cross channel projection from the retarded $D$-channel vertex to the $C$- and $P$-channels leads to the transfer frequency being fixed to the internal frequency of the loops, i.e., $it_0 = \pm i\Lambda$.
Effectively the projected full vertices $V^{\Lambda, X}$ in \cref{eq:frg-compact-P,eq:frg-compact-C,eq:frg-compact-D} are replaced by
\begin{align}
    \label{eq:static-vertex-mod-P}
    V^{\Lambda, P} &{}\rightarrow V^{S,\Lambda, P} + \mathbb P\big[V^{R,\Lambda}(it_0 = i\Lambda)\big] \,,\\
    \label{eq:static-vertex-mod-C}
    V^{\Lambda, C} &{}\rightarrow V^{S,\Lambda, C} + \mathbb C\big[V^{R,\Lambda}(it_0 = i\Lambda)\big] \,,\\
    \label{eq:static-vertex-mod-D}
    V^{\Lambda, D} &{}\rightarrow V^{S,\Lambda, D} + \mathbb D\big[V^{R,\Lambda}(it_0 = 0) \big] \,,
\end{align}
where $\mathbb X\big[V^{R,\Lambda}(it_0)\big]$ is the (real/momentum space)
projection of the retarded vertex (written in the $D$-channel)
into the $X$-channel.
This modification to the static flow equations is visualized as colored wiggly lines in the diagrams of \cref{fig:flow_diagrams}.

Ignoring cross-channel projections, i.e., effectively performing RPA in each of the diagrammatic channels~\cite{Fischer2024s}, the retarded interaction naturally favors charge bond-order formation provided the particle-hole susceptibility is of the appropriate form (see \cref{sec:dchanfrg}).
However, when allowing for feedback from the $D$-channel to the other two channels and vice-versa, interesting interplay between the different fluctuation types arises, which can lead to both cooperation and competition of different phases (which of course includes the case of conventional superconductivity).
Notably, the effect of the retardation is included in the electron-phonon vertex contributions as factor $1/\omega_{\nu,\bv q}^2$ ($D$-channel) and $1/(\omega_{\nu,\bv q}^2 + \Lambda^2)$ ($P$- \& $C$-channels).
This leads to an effective screening of electronic corrections ($P$- \& $C$-channels) for scales $\Lambda \gg \omega_{\nu ,\bv{q}}$, while for $\Lambda \lesssim \omega_{\nu ,\bv{q}}$ all terms contribute roughly equally.
Such treatment of retardation stands in contrast to some of the earlier studies~\cite{Classen2014i} of electron-phonon coupling in the electronic FRG, where retardation effects have been ignored.

\subsection{Truncated Unity FRG}\label{ssec:TUFRG}
Having discussed the frequency dependencies and the effective static flow equations, we now turn to the numerical approximations employed to solve them. In the context of quantum materials, the truncated unity (TU) approximation to FRG~\cite{Husemann2009e, Lichtenstein2017, Beyer2022r, Profe2022t} has emerged as highly efficient formulation allowing to treat multi-orbital, non-$SU(2)$ and three-dimensional systems~\cite{Platt2013, Kiesel2013u, Wang2013c,
Delapena2019c, Klebl2022m, Profe2024k, Profe2024m, Fischer2025t, Beck2025k, Durrnagel2025a, Bigi2025p, Guo2025a}, to the extent of a community code that interfaces to major \emph{ab initio} software packages~\cite{10.21468/SciPostPhysCodeb.26, 10.21468/SciPostPhysCodeb.26-r0.5}. It makes use of the vertex functions' channel decomposition \cref{eq:frg-channels} by approximating fermionic momentum dependencies as truncated lattice Fourier transform.
Physically, this truncation corresponds to limited range real-space bonds of electronic bilinears. In static approximation, the FRG flow equations take the following form:
\begin{align}
    \label{eq:tufrg-P}
    \dot P^\Lambda_{mn}(\bv s) &{}= \sum_{ij} V^{\Lambda, P}_{mi}(\bv s) L^{\Lambda, P}_{ij}(\bv s) V^{\Lambda, P}_{jn}(\bv s) \,, \\
    \label{eq:tufrg-C}
    \dot C^\Lambda_{mn}(\bv u) &{}= \sum_{ij} V^{\Lambda, C}_{mi}(\bv u) L^{\Lambda, H}_{ij}(\bv u) V^{\Lambda, C}_{jn}(\bv u) \,, \\
    \label{eq:tufrg-D}
    \dot D^\Lambda_{mn}(\bv t) &{}= \begin{multlined}[t]
    -2\sum_{ij} \bigg( V^{\Lambda, D}_{mi}(\bv t) - \frac12 V^{\Lambda,C}_{mi}(\bv t)\bigg) L^{\Lambda,H}_{ij}(\bv t) \\
    \bigg( V^{\Lambda, D}_{jn}(\bv t) - \frac12 V^{\Lambda,C}_{jn}(\bv t)\bigg)
    + \frac12\dot C^\Lambda_{mn}(\bv t) \,.
    \end{multlined}
\end{align}
$L^{\Lambda,P/H}$ are integrated scale derivatives of particle-particle and particle-hole loops, respectively [cf.~\cref{eq:loopdev}] and read
\begin{multline}
    \label{eq:loopint}
    L^{\Lambda,P/H}_{ij}(\bv q) = \frac1{2\pi N_{\bv k}} \sum_{\bv k} \Big[ G(i\Lambda, \bv k) G(\pm i\Lambda, \pm \bv q \mp \bv k) \\ {} + G(-i\Lambda, \bv k) G(\mp i\Lambda, \pm \bv q \mp\bv k) \Big] \, f_i^*(\bv k) f_j^{\phantom{*}}(\bv k) \,,
\end{multline}
where $f_i(\bv k)$ is the formfactor for the bond $i$.\footnote{%
We note that not only for the treatment of retardation the sharp frequency cutoff turns out to be crucial, but also for the numerical simplification of \cref{eq:loopint}: The structure allows to employ fast Fourier transformations to calculate the convolution, which renders the numerical complexity $\mathcal O(N_{\bv k} \log N_{\bv k})$.%
}

Repeated scattering in a certain channel can lead to a divergence of one or more eigenvalues of the two-particle vertex at some RG scale $\Lambda_c$.
In such a case, we consider the effective static interactions in the \emph{physical interaction channels}:
\begin{equation}
    \Phi^\Lambda_{\mathrm{charge}} = 2 V^{\Lambda, D} - V^{\Lambda, C}
    ,~
    \Phi^\Lambda_{\mathrm{spin}} = -V^{\Lambda, C}
    ,~
    \Phi^\Lambda_{\mathrm{SC}} = V^{\Lambda, P}.
    \label{eq:physical_channels}
\end{equation}
The divergent (negative) eigenvalue indicates a phase transition to a charge/spin/superconducting (SC) order at a temperature on the order of $\Lambda_c$. The corresponding eigenvector allows us to extract the dominant exchange momentum of the order and its real space profile---distinguishing between, e.g., $s$-wave and $d$-wave superconductivity.

In practice, we implement the effect of retarded density-density interactions as detailed in \crefrange{ssec:FRG}{ssec:TUFRG} on top of the TUFRG backend of the \textsc{divERGe} library~\cite{10.21468/SciPostPhysCodeb.26, 10.21468/SciPostPhysCodeb.26-r0.5}---we need only modify the content of vertices according to \crefrange{eq:static-vertex-mod-P}{eq:static-vertex-mod-D} during the iteration of the RG differential equation.\footnote{An extension to non-interacting bosons in either of the diagrammatic channels is trivial, as the procedure is analogous no matter which channel the bosons are native to---electron-plasmon, electron-magnon, or electron-cooperon coupled systems could be studied with the exact same approach given the energy scales separate accordingly.}

\subsection{Phononic softening}
\begin{figure}
    \centering
    \includegraphics{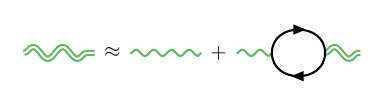}
    \caption{The lowest order correction to the phonon propagator amounts to an RPA-like series through the Phonon Dyson \cref{eq:phonon_Dyson}.}
    \label{fig:phonon_self_energy}
\end{figure}

Apart from the electronic picture described by FRG, the onset of charge order in a coupled electron-phonon system can also be determined by the freezing of a soft phonon mode~\cite{Nomura2015, Hall2019, Berges2020, Giustino2017}.
This can be described by the phonon Dyson equation (in band space):
\begin{multline}
    \tilde G^\mathrm{ph}_{\bv{q}\mu\nu}(i\omega) = G^\mathrm{ph}_{\bv{q}\mu}(i\omega)\, \delta_{\mu\nu} + G^\mathrm{ph}_{\bv{q}\mu}(i\omega) \\
    (2\omega_{\mu,\bv q})^{-1}\,\Pi_{\bv{q}\mu \nu}(i\omega) \, \tilde G^\mathrm{ph}_{\bv{q}\nu}(i\omega) \,,
    \label{eq:phonon_Dyson}
\end{multline}
with $\tilde G^\mathrm{ph}$ the renormalized phonon propagator and $\Pi$ the phonon self-energy.
The solution of the Dyson \cref{eq:phonon_Dyson} is the dressed phonon-propagator
\begin{equation}
    \tilde G^\mathrm{ph}_{\bv{q}}(i\omega) = \frac{G^\mathrm{ph}_{\bv{q}}(i\omega)}{1- G^\mathrm{ph}_{\bv{q}}(i\omega) \, (2\omega_{\bv q})^{-1}\,\Pi_{\bv{q}}(i\omega)} \,.
    \label{eq:dressed_phonon_propagator}
\end{equation}
The renormalized phonon frequency is obtained from considering the poles of \cref{eq:dressed_phonon_propagator} with respect to $i\omega$, which amounts to diagonalizing the dressed frequencies $\tilde\omega$ (in static approximation):
\begin{equation}
    \tilde\omega^2_{T;\bv{q} \mu\nu} = \omega^2_{\nu,\bv{q}} \, \delta_{\mu \nu} +  \Pi_{T;\bv{q}\mu \nu}(i\omega=0) \,.
    \label{eq:ph_renormalization}
\end{equation}
In perturbation theory, the leading contribution for the phonon self-energy reads
\begin{equation}
    \Pi_{T;\bv{q} \mu \nu} = \frac{2}{N_{\bv k}}\sum_{\bv{k}} \bar g^{(\mu)}_{\bv{q}, \bv{k}}\, \chi_{T;\bv{q},\bv{k}}^{H}\, g^{(\nu)}_{\bv{q},\bv{k}} \,,
    \label{eq:ph_self_energy}
\end{equation}
with the factor $2$ due to $SU(2)$ symmetry and $\chi_{T;\bv q,\bv k}^{H}$ the electronic particle-hole susceptibility at a temperature $T$ (analogous to \cref{eq:loopdev}, replacing scale with temperature dependence). Diagrammatically, the insertion of \cref{eq:ph_self_energy} into the Dyson \cref{eq:phonon_Dyson} corresponds to an RPA-like resummation of electronic screening processes, cf.~\cref{fig:phonon_self_energy}.
Upon lowering the temperature, the magnitude of $\chi_{T;\bv{q}, \bv{k}}^{H} < 0$ increases, possibly reaching the point where $\tilde \omega^2_{T;\bv q\mu \nu}$ has a negative eigenvalue for some $\bv{q}$.
This indicates a structural transition that is equivalent to a charge (bond) order in the electronic sector.
In one dimension the particle-hole bubble diverges already at the bare level guaranteeing a Peierls transition at some temperature $T_c$, while in two or three dimensions the situation lacks such generality.

\subsection{FRG restricted to the {\itshape D}-channel}
\label{sec:dchanfrg}
It is instructive to draw the analogy of the leading phonon self-energy correction to a fully electronic picture. As suggested by the diagrams in \cref{fig:phonon_self_energy}, the two approaches should be fully equivalent when performing 
an RPA resummation of charge fluctuations in the electronic picture ($D$-RPA).
In practical calculations, we make use of the fact that---within the scope of the approximations taken---the FRG restricted to the $D$-channel amounts to precisely the $D$-RPA; except for slight variations in temperature/scale dependence~\cite{Fischer2024s}.

Omitting the $P$- \& $C$-channels, the flow equation for the renormalization of the vertex in the $D$-channel \cref{eq:tufrg-D} simplifies to
\begin{equation}
    \label{eq:flow_equation_D_channel}
    \dot D^\Lambda_{mn}(\bv t) = -2\sum_{ij} V^{\Lambda,D}_{mi}(\bv t) L^{\Lambda, H}_{ij}(\bv t) V^{\Lambda,D}_{jn}(\bv t) \,,
\end{equation}
where $V^{\Lambda,D} = D^\Lambda + \mathbb{D}[V^{R,\Lambda}(it_0=0)]$, cf.~\cref{eq:static-vertex-mod-D}. When considering the case of phonon-induced interaction only, $D^\Lambda$ vanishes on the bare level and we can take $D^\infty=\mathbb D[V^{R,\Lambda}(it_0=0)]$ as initial condition instead of the replacement \cref{eq:static-vertex-mod-D}. It is straightforward to show that \cref{eq:flow_equation_D_channel} is then solved by
\begin{equation}
    D^\Lambda(\bv t) = \frac{D^\infty(\bv t)}{1-2D^\infty(\bv t) \int_{\infty}^{\Lambda}\dd\Lambda'\, L^{\Lambda',H}(\bv t)} \,.
    \label{eq:D_channel_RPA}
\end{equation}
Writing the solution in terms of phonon propagators, one notices that it corresponds to dressing the bare phonon propagator by its leading order correction \cref{fig:phonon_self_energy}, with the particle-hole bubble replaced by a scale integration over the (integrated) loop derivative
\begin{equation}
    \chi^{H}_{\bv t} \to \int_{\infty}^{\Lambda}\dd\Lambda' \, L^{\Lambda',H}(\bv t)\,.
    \label{eq:looprepl}
\end{equation}
We note that the dependency on temperature in \cref{eq:ph_self_energy} is replaced by one on a minimal scale $\Lambda$ that the integral \cref{eq:looprepl} extends to.

In contrast, the full FRG additionally contains effective renormalization of the electron-phonon coupling via higher order cross-channel vertex corrections, as well as the self-energy contribution from the $D$-RPA diagrams.
The former are particularly severe in---but not restricted to---the case of bare-level electron-electron interaction, where the electron-phonon and electron-electron interactions intertwine diagrammatically.

\subsection{Usage in combination with \emph{ab initio} calculations}

In practical scenarios beyond the most simplified lattice models, the phonon bands and EPCs are typically calculated via \emph{ab initio} methods such as density-functional perturbation theory (DFPT)~\cite{Gonze1997r, Baroni2001p, Giustino2017}. This method in particular already takes electronic screening of both EPC and phonon bands into account, and is able to describe RPA-type lattice instabilities.
For prior work demonstrating the practical equivalence of the calculation in \cref{eq:ph_renormalization} and DFPT screening we refer to Refs.~\cite{Nomura2015, Berges2020b}.
The diagrammatic corrections to the phonon frequencies in DFPT can be expressed as
\begin{equation}
    \label{eq:ph_renormalization_DFPT}
    \tilde\omega^2_{\bv{q} \mu\nu} = \omega^2_{\nu, \bv{q}} \delta_{\mu \nu} + \underbrace{ \frac{2}{N_k}\sum_{\bv{k}} \bar g^{(\mu)}_{\bv{q}, \bv{k}}\,  \chi_{0;\bv{q},\bv{k}}^{H}\, g^{\mathrm{scr}\,(\nu)}_{\bv{q},\bv{k}} }_{\displaystyle \Pi^\mathrm{DFPT}_{\bv q\mu\nu}} \,,
\end{equation}
where screening of the electron-phonon coupling is accounted for in $g^\mathrm{scr}$ and $\chi_{0;\bv{q},\bv{k}}^{H}$ is the particle-hole susceptibility at zero temperature.
To avoid double counting 
of low-energy states in multi particle calculations, one needs to either unscreen both phonon frequencies and electron-phonon coupling up to some energy scale $\Lambda_0$ or work with constrained DFPT (cDFPT)~\cite{Nomura2015}, where screening contributions from the low-energy subspace are explicitly removed.
This allows for a subsequent unbiased treatment of the electronic fluctuation channels in the remaining energy window via FRG in a rigorous manner.

\section{Results} \label{sec:results}
After setting up the general scheme for applying the all-electronic FRG to arbitrary electron-phonon models in the previous section, we now exemplify this machinery for the square lattice Hubbard model with acoustic phonons derived in \cref{sec:model}. By doing so, we not only explicitly demonstrate the equivalence between phonon softening described by \cref{eq:ph_renormalization} and the renormalization of the effective vertex in $D$-RPA \cref{eq:D_channel_RPA} in the absence of electronic correlations, but we also investigate how this picture alters upon including screening processes from all diagrammatic channels within the FRG.
Finally, by taking into account both electron-phonon and electron-electron interactions on equal footing within our unbiased FRG treatment we discuss the qualitative features of the such obtained phase diagram in relation to previous studies of the all-electronic Hubbard model.

As both the phonon dispersion and the electron-phonon coupling are momentum dependent, we track their respective strengths for simplicity via
\begin{equation}
\begin{aligned}
    \lambda &{}\equiv \frac{g^2}{W M \omega^2} \,,\\
    g &{}\equiv \ g_{x/y, (\pi,\pi)} = 2|\beta t| \,,\\
    \omega^2 &{}\equiv \omega_{x/y, (\pi,\pi)}^2 = 4\left(K_\parallel + K_\perp\right) \,,\\
    \alpha^2 &{}\equiv \frac{\omega^2_{x, (\pi,0)}}{\omega^2_{x/y, (\pi,\pi)}} = \frac{K_\parallel}{K_\parallel + K_\perp} \,,
\end{aligned}
\label{eq:parameters}
\end{equation}
where $W = 8t$ is the electronic band width and the indices $\{\nu = x/y,\ \bv q = (\pi,\pi)\}$ indicate the two breathing modes at the $M$-point and $\{\nu = x,\ \bv q =(\pi,0)\}$ the breathing mode at the $X$-point.
Since the shear mode $\{\nu = y,\ \bv q =(\pi,0)\}$ does not carry EPC [cf.~\cref{fig:model}(d)], a variation of $K_\perp$ allows for a tuning of the phonon band structure, and thus the retarded vertex \cref{eq:retarded_intn}, without changing the structure of the coupling.

\subsection{Peierls instability: phonon softening \& {\itshape D}-RPA}

\subsubsection{Bond orders in the phononic picture}
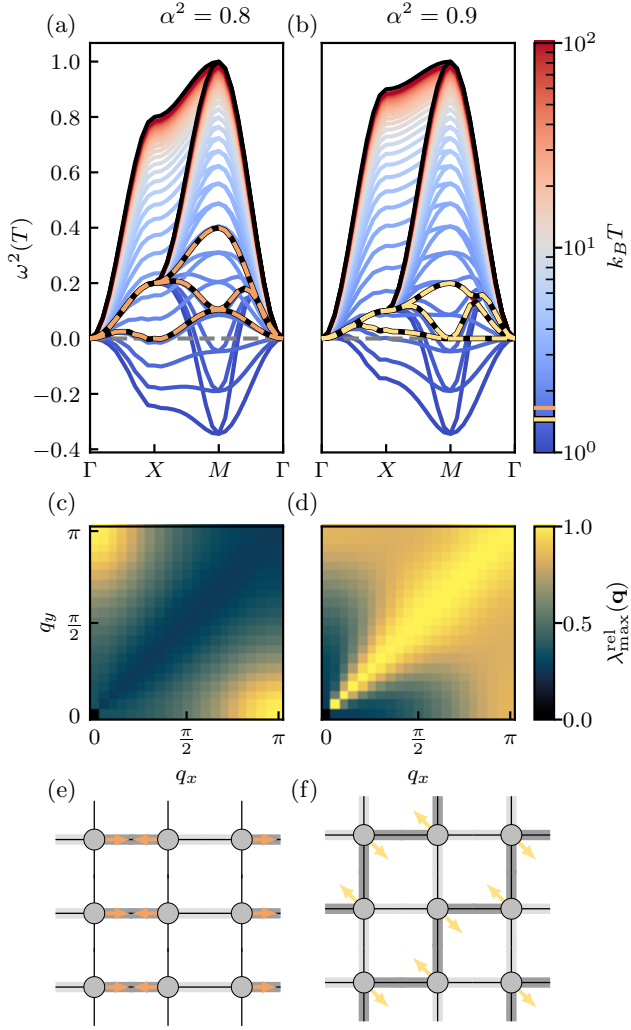
\begin{figure}
    \centering
    \vspace{-0pt}
    \begin{tikzpicture}[inner sep=0pt, outer sep=0pt]
    \node (a) at (0,0) {
        \includegraphics[width=1.0\columnwidth]{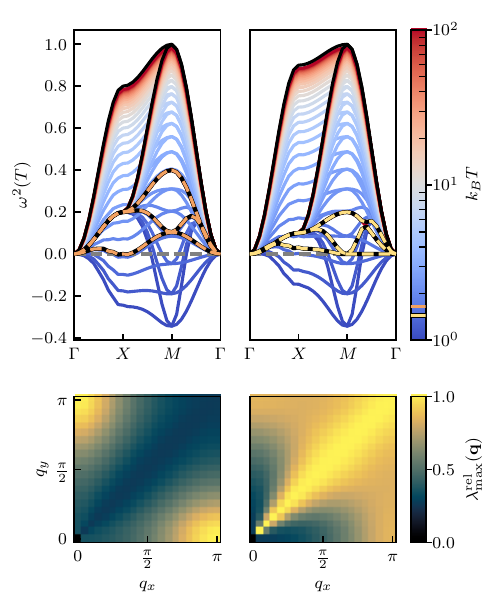}
    };
    \node[anchor=north, xshift = -6, yshift = 8pt] at (a.south){
        \includegraphics[width=0.4\columnwidth]{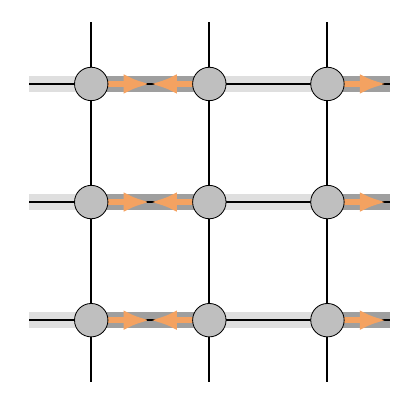}
        \includegraphics[width=0.4\columnwidth]{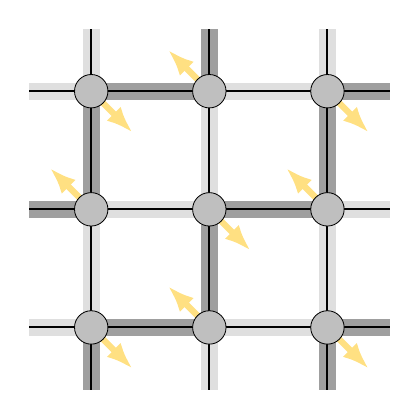}
    };
    \node[anchor=south east] at (-3.2,4.8) {\small(a)};
    \node[anchor=south east] at (0,4.8) {\small(b)};
    \node[anchor=north east] at (-3.2,-1.2) {\small(c)};
    \node[anchor=north east] at (   0,-1.2) {\small(d)};
    \node[anchor=south] at (-1.5,5.0) {$\alpha^2 = 0.8$};
    \node[anchor=south] at ( 1.5,5.0) {$\alpha^2 = 0.9$};
    \node[anchor=north east] at (-3.2,-5) {\small(e)};
    \node[anchor=north east] at (   0,-5) {\small(f)};
    \end{tikzpicture}
    \vspace{-15pt}
    \caption{
    (a,b): Renormalization of the phonon bands according to \cref{eq:ph_renormalization} for $\lambda = 0.4$, $\omega = 1$ and $\alpha^2 \in \{0.8, 0.9\}$ (left to right) as the temperature is decreased.
    Renormalized phonon bands at the critical temperatures $k_\mathrm{B} T_c/t \approx 1.65\ (1.45)$ are given in orange (yellow).
    For low $\alpha$ there is an instability at $\bv{q} = X$ and for high $\alpha$ we have a simultaneous instability on the line between $\Gamma$ and $M$.
    (c,d): Normalized leading eigenvalues for each $\bv q$ of the effective electron-electron interaction for the same parameters and $U = 0$ in the $D$-RPA.
    The critical scales in the RPA $\Lambda_c/t = 3.31\ (2.83)$ have a roughly linear relationship to the critical temperatures here.
    Panels~(e,f) show the real-space structure of both the frozen phonon modes (lattice relaxations are indicated by arrows) and corresponding charge bond order (enhanced/reduced hoppings are highlighted by dark/light gray shading) at $q=X$ (e) and $q=M$ ($\sigma_{xy}$-odd branch) (f).
    }
    \label{fig:renormalized_phonon_bands}
\end{figure}

For $U = 0$, we can obtain the lattice instability from \cref{eq:ph_renormalization} by successively lowering the temperature of the system and inspecting the momentum at which the phonon dispersion first becomes imaginary. The precise shape of the structural instability can be deduced from the associated phonon mode.
Applying this to the model of \cref{sec:model}, one obtains renormalized phonon bands as a function of temperature shown in \cref{fig:renormalized_phonon_bands}(a,b).
Notably, this differs from the usual workflow of DFPT calculations, where the renormalized phonon dispersion is calculated at $T \sim 0$ with some fixed smearing. Here instead, the electronic screening is only taken into account up to $T_c$, where the perturbative approximation breaks down.

By inspecting the phonon dispersion at the critical temperature, a competition between phonon modes with variable exchange momenta $\bv q=(q_x,0)$ ($X$-type) and $\bv q= (q_x,q_x)$ ($M$-type) becomes apparent.
While the $X$-type order has a well defined $q\in \overline{\Gamma X}$ in \cref{fig:renormalized_phonon_bands}(a) and is degenerate with the symmetry equivalent point on the $\overline{\Gamma Y}$ line, all $q\in \overline{\Gamma M}$ in the $M$-type order are degenerate [cf.~\cref{fig:renormalized_phonon_bands}(b)].
To ascertain the dominant superposition at this point, one has to minimize the free energy in the symmetry broken phase within the eigenspace spanned by the degenerate instabilities by a subsequent mean-field (or Ginzburg-Landau) treatment.
For simplicity however, we will only consider one of the degenerate states in the following and we focus on the mechanism for the competition of the two charge-bond orders which is revealed by considering \cref{eq:ph_renormalization}:

For relatively small $\omega(\bv{q}=X) \propto \alpha$, the elevated temperature scale renders electronic nesting less important and the phononic self-energy $\Pi$ is almost uniform in $\bv q$. In this regime, the lower bare frequency $\omega_{x,X}<\omega_{x,M}$ is enough to favor the $X$-type order as depicted in \cref{fig:renormalized_phonon_bands}(a).
In the converse regime, electronic correlations eventually dominate to drive the $M$-type order: at low temperatures nesting of the particle-hole bubble $\chi^H(\bv q)$ becomes relevant and is strongly peaked at $\bv q=M$ resulting in a large peak of the phonon self-energy in \cref{eq:ph_self_energy}.
In fact, both $\Pi_{\bv{q}\mu\nu}$ and $\omega^2_{\bv{q}\nu}$ are proportional for the relevant mode on the $\overline{\Gamma M}$ line, resulting in the simultaneous freezing of the renormalized phonons on this whole line as attested by \cref{fig:renormalized_phonon_bands}(b).

The eigenvectors at the $X$ and $M$ point reveal the lattice relaxations associated with the frozen phonon modes and are indicated by arrows in \cref{fig:renormalized_phonon_bands}(e,f):
At the $X$-point, the phonon instability has a $p_x$-like stripe structure where elongated and squeezed nearest-neighbor bonds alternate in the $X$-direction.
In turn, the phonon dispersion is doubly degenerate at the $M$ point and the eigenspace is spanned by the same $p_x$-like distortion encountered at the $X$-point and its $C_4$ rotated $p_y$-like partner.
For any other point on the $\overline{\Gamma M}$ line however, this degeneracy is lifted and the $\sigma_{xy}$ mirror odd mode freezes ($x\leftrightarrow y$).
Hence we depict in \cref{fig:renormalized_phonon_bands}(f) the odd superposition $p_x - p_y$.
We note that these two orders are simply the square-lattice equivalents of the Peierls instability of the linear chain. Unlike in the 1D Peierls case, the particle-hole bubble diverges only logarithmically for the nesting momentum $\bv q = M$, so that the latter is not necessarily dominant at intermediate coupling.

\subsubsection{Electronic perspective: {\itshape D}-RPA}
As established in \cref{sec:dchanfrg}, the renormalization of the phonon self-energy is formally equivalent with an FRG treatment of the phonon-mediated effective electron-electron interaction restricted to the $D$-channel.
Comparing the leading eigenvalues of the effective $D$-RPA vertex at the critical scale with the frozen phonon modes at $T_c$ in \cref{fig:renormalized_phonon_bands}, we indeed find a transition as a function of $\alpha$ that is consistent between both formalisms.

From the all-electronic perspective of the $D$-RPA, the low $\alpha$ implies elevated critical scales $\Lambda_c$, where the electronic renormalization does not alter the qualitative features of the effective vertex significantly.
Hence, the dominant instability is already encoded in the bare phonon-mediated electron-electron vertex $D^{\Lambda_c}(\bv{q}) \sim D^\infty(\bv{q})$ in \cref{eq:D_channel_RPA} and peaks at the $X$ point [compare \cref{fig:V_eff,fig:renormalized_phonon_bands}(c)].
An increase in $\alpha=\omega(\bv{q} = X)/\omega(\bv{q} = M)$ suppresses charge fluctuations at $\bv q = X$ on the bare level, i.e., $D^\infty(\bv q=X)$, and consequently lowers the critical scale of this transition.
In this scenario, the momentum structure of the particle-hole bubble can enter more prominently in the vertex renormalization and the $M$-type CBO is again favored.
The eigenvalues relevant for the $M$-type charge bond order are exactly degenerate on the $\overline{\Gamma M}$ line [cf.~\cref{fig:renormalized_phonon_bands}(d)], which corresponds to the freezing of a phonon band with unspecified momentum along the $\overline{\Gamma M}$ line as discussed earlier.
Directly at the $M$ point, the leading eigenvector is doubly degenerate.
This is consistent with the doubly degenerate phonon mode at $M$ in \cref{fig:renormalized_phonon_bands}(b).

In the all-electronic formalism, lattice relaxation and the concomitant squeezing and stretching of bonds surfaces as enhanced and reduced hybridization elements since the overlap of Wannier orbitals scales with the inter-atomic distance. Therefore, the Peierls-like instabilities in the phononic picture correspond to  charge bond orders (CBOs) in the electronic sector, where the hopping amplitude alternates along the 1D chains.
Comparing the CBO order parameter with the lattice relaxation of the frozen phonon modes in \cref{fig:renormalized_phonon_bands}(e,f) indeed reveals a consistent picture not only for the ordering wavevector $\bv q$ but also the detailed shape of the instability.

\subsubsection{$D$-RPA phase diagram}
After having confirmed the equivalence of the phonon softening mechanism and $D$-RPA, we can investigate the full ($\lambda$, $\alpha^2$) phase space.
From now on, we shall classify the phases by their dominant phase type (e.g.~CBO), momentum $\bv q_c$, and irreducible representation (irrep, $\mathrm{I}$) of the local point group, which amounts to $\mathrm{CBO}_{\bv q_c}^\mathrm{I}$.
The little group at, for example, the $X$ point is given by $C_\text{2v}$ and the $p_x$-type order of \cref{fig:renormalized_phonon_bands}(e) transforms in its $B_1$ irrep.
In turn, the $M$ point features full $C_{4v}$ symmetry and the order parameter transforms can be associated with the two dimensional $E$ irrep.

\begin{figure*}[!t]
    \centering
    \vspace{-15pt}
    \begin{tikzpicture}[inner sep=0pt, outer sep=0pt]
    \node (a) at (0,0) {
        \includegraphics[width=\linewidth]{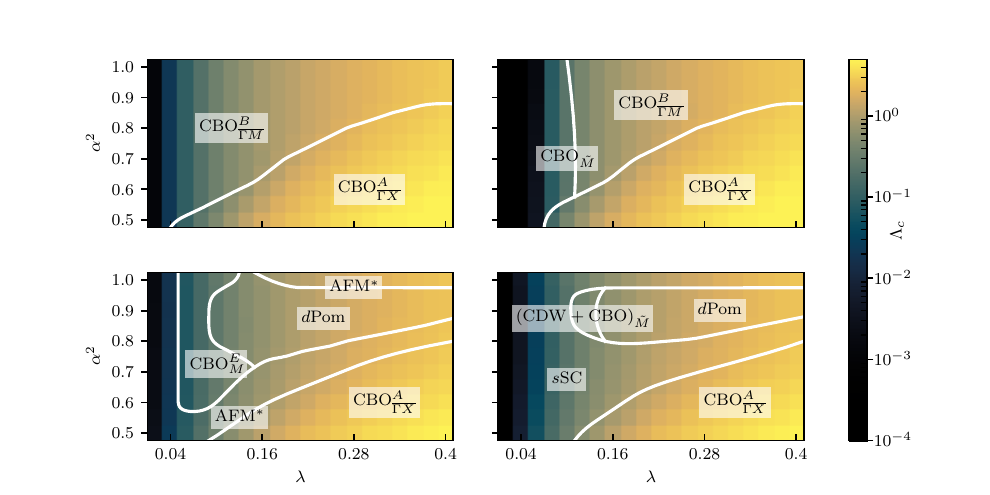}
    };
    \node[anchor=south west, xshift= 70pt, yshift=-25pt] at (a.north west) {\small(a)};
    \node[anchor=south west, xshift=  0pt, yshift=-25pt] at (a.north) {\small(b)};
    \node[anchor=north west, xshift= 70pt, yshift=  0pt] at (a.west) {\small(c)};
    \node[anchor=north west, xshift=  0pt, yshift=  0pt] at (a) {\small(d)};
    \node[anchor=south east, xshift=-80pt, yshift=-20pt] at (a.north) {$n = 1.0$};
    \node[anchor=south west, xshift= 65pt, yshift=-20pt] at (a.north) {$n = 0.85$};
    \node[anchor=east, xshift=50pt, yshift=-55pt] at (a.north west) {D-RPA};
    \node[anchor=east, xshift=50pt, yshift=90pt] at (a.south west) {FRG};
    \end{tikzpicture}
    \vspace{-20pt}
    \caption{Phase diagram in 
    ($\lambda, \alpha^2$) space with $U=0$, $\omega = 0.1$ at half filling $\langle n \rangle = 1.0$ (left) and slightly away $\langle n \rangle = 0.85$ (right).
    The top row (a,b) shows the $D$-RPA and the bottom row (c,d) the full FRG result.
    Colors indicate the critical scale of the structural instability.
    The nomenclature for the different phases is explained in the text.
    }
    \label{fig:lambda_alpha_phase_diagram}
\end{figure*}
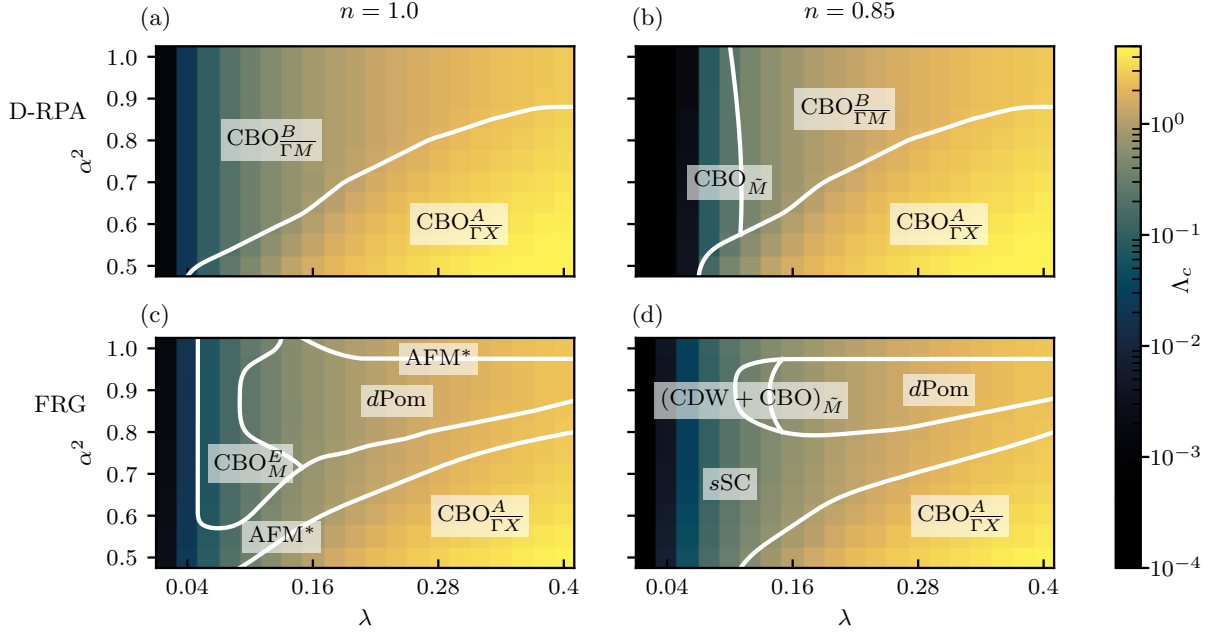

In \cref{fig:lambda_alpha_phase_diagram}(a), we depict the phase diagram at half filling, confirming the intuition extracted from \cref{fig:renormalized_phonon_bands}:
Since an enhancement of the electron-phonon coupling $\lambda$ directly transfers to an enhanced critical temperature/scale, an increase of $\lambda$ has a similar effect as a reduction of $\alpha$.
Consequently, the $X$-type CBO ($\mathrm{CBO}_{\overline{\Gamma X}}^\mathrm{A}$) is situated at large $\lambda$ and low $\alpha$.
While deep in this regime, the $\bv q$ vector of the instability is located directly at the $X$ point [as in \cref{fig:renormalized_phonon_bands}(a,c)]. There is a small regime in the phase diagram where $q_x$ decreases in a continuous manner as either $\alpha$ is increased or $\lambda$ decreased.
More precisely, the $X$-type order on the $\overline{\Gamma X}$ line is always even under $\sigma_y$ that on the $X$ point becomes the $B_1$ irrep when additional $\sigma_x$ is present.
At even larger $\alpha^2 / \lambda$, the system eventually transitions to the $M$-type CBO ($\mathrm{CBO}_{\overline{\Gamma M}}^\mathrm{B}$) phase.
These $M$-type states, however, are bound to feature a degeneracy along the $\overline{\Gamma M}$ line for all model parameters by the underlying symmetry discussed above.
However, all states are odd under the remaining $\sigma_{xy}$ mirror, including the $\Gamma$ point ($B_1$ irrep) and the $M$ point, where it corresponds to a superposition $p_x - p_y$ in the $E$ irrep (and thus is degenerate with $p_x+p_y$).

Upon introducing moderate hole doping in \cref{fig:lambda_alpha_phase_diagram}(b), the general structure of the $D$-RPA phase diagram remains intact.
However, since the doping perturbs the exact nesting at the $M$ point, there is a transition for low $\lambda$ from $q=(q_x,q_x)$ order to incommensurate momenta close to the $M$ point, that we term $\tilde M$-type ($\mathrm{CBO}_{\tilde M}$) order.
Since in general $\tilde M$ does not lie on a high-symmetry path it does not feature local symmetries and thus cannot be classified further by irreducible representations.

\subsection{Functional renormalization group}
\subsubsection{Charge orders}

When turning on the $P$ and $C$-channels in the FRG and allowing for cross channel projections, the general structure of the $D$-RPA phase diagram is maintained:
The $X$-type CBO phase, that is primarily governed by the bare vertex due to large $\Lambda_{c}$, persists and is only pushed to lower $\alpha^2$ values, since the FRG incorporates more screening effects generically lowering the critical scales.
The remaining phase space is still governed by instabilities on the $\overline{\Gamma M}$ line due to the dominant Fermi surface (FS) nesting at $\bv q = M$, but the massive degeneracy on this line is lifted by cross-channel projections and pins the condensation momentum to one of the high symmetry points.

Inside the large $\alpha^2$ regime, we obtain a competition between three types of orders: The $p_{x/y}$-type CBO at $\bv q=M$ [$\mathrm{CBO}_{M}^\mathrm{E}$ depicted in \cref{fig:renormalized_phonon_bands}(f)] surfaces as dominant instability at moderate $\lambda$, while for larger $\lambda$ the dominant momentum converges to $\bv q \rightarrow 0$ as momentum grid resolution is increased.
Strictly speaking, the transfer momentum is never identically zero, because we have to fix $V^{(\nu=0)}(\bv q=0)=0$ in order to avoid discontinuities from the different directional limits while keeping all symmetries intact (see \cref{ssec:retarded_interaction}).
Asymptotically, we may still identify this instability with a $d$-wave Pomeranchuk ($d\mathrm{Pom}$) order
$\sum_{\sigma \in \{\uparrow, \downarrow\}}
\braket{c^\dagger_{j+\delta\sigma} c_{j\sigma}^{\vphantom{\dagger}}} \sim \left(\delta_{\delta \pm\hat x} - \delta_{\delta \pm\hat y}\right)$, which breaks rotational and retains translational symmetry.

While these phases are already present in the degenerate manifold of the $D$-RPA diagram, the FRG treatment results in an additional phase characterized by a degenerate antiferromagnetic ($\mathrm{AFM}_M$), charge-density-wave ($\mathrm{CDW}_M$) and s-wave superconducting ($s\mathrm{SC}$) instability (termed $\mathrm{AFM}^*$).
The origin of this degeneracy lies in the $O(4)$ symmetry of the system in the absence of Hubbard interaction and at half filling and has previously been discussed in the context of non-dispersive, optical phonons~\cite{Götz2022, Yang2022}.
In the $D$-RPA, this symmetry is explicitly broken by the biased resummation leading to an absence of these phases in \cref{fig:lambda_alpha_phase_diagram}(a).

\begin{figure}
    \centering
    \vspace{-0pt}
    \begin{tikzpicture}[inner sep=0pt, outer sep=0pt]
    \node (a) at (0,0) {
        \includegraphics[width=1.0\linewidth]{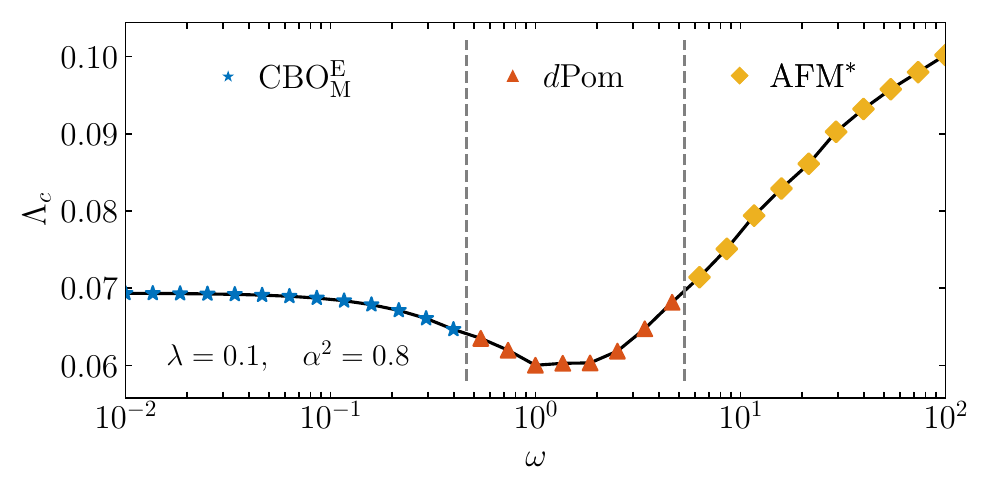}
    };
    \node[anchor=north east, xshift =-30, yshift = 35pt] (a1) at (a.center){
        \includegraphics[width=0.15\columnwidth]{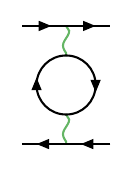}
    };
    \node[anchor=north, xshift = 20, yshift = 30pt] (a2) at (a.center){
        \includegraphics[width=0.2\columnwidth]{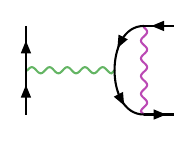}
    };
    \node[anchor=north west, xshift = 62, yshift = 6pt] (a3) at (a.center){
        \includegraphics[width=0.18\columnwidth]{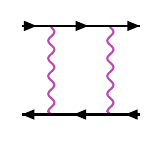}
    };
    \end{tikzpicture}
    \vspace{-20pt}
    \caption{
    Critical scale $\Lambda_c$ and leading instability as a function of $\omega$ for $U=0$, $\lambda = 0.1$, $\alpha^2 = 0.8$.
    The diagrams visualize the dominant perturbative contributions at one-loop level to the effective vertex associated with the different phases.
    The $C$-channel vertex (purple wiggly line) is suppressed for low $\omega$ due to retardation, promoting a dominant role of $D$-channel (green wiggly line) in this regime.
    }
    \label{fig:omega_sweep}
\end{figure}

To understand the parameter dependence of the dominant symmetry breaking propensity, it is instructive to investigate the dependence of electronic screening processes on the retarded electron-phonon interaction as a function of $\omega$, that is related to the phase space of \cref{fig:lambda_alpha_phase_diagram} via \cref{eq:parameters}.
The channel selective retardation of \cref{eq:frg-compact-P,eq:frg-compact-C,eq:frg-compact-D} establishes a hierarchy between the one-loop diagrams in the FRG flow equation~\cref{fig:flow_diagrams}:
For low $\omega \ll \Lambda_c$, the retarded interaction components in the $C$-channel (purple lines) are weakened and only $D$-channel vertices (green lines) are relevant at the bare level, which through cross-channel projection into the $C$-channel break the degeneracy along $\overline{\Gamma M}$ in favor of CBO at the $M$ point.
As $\omega$ is increased ($\omega \gtrsim \Lambda_c$), the mixed diagrams start to contribute and suppress the CBO and promote the $d\mathrm{Pom}$ order.
In the large $\omega \gg \Lambda_c$ limit, the hierarchy between the channels is lifted as the interaction is effectively instantaneous.
There, antiferromagnetic fluctuations in the $C$-channel are strong, yielding the degenerate AFM$^*$ phase.\footnote{
Nontrivial degeneracies among different two-particle channels have been recently discussed in the context of pseudospin symmetries in Hubbard-like models in Ref.~\cite{meixner2026degeneracies}. The unbiased nature of FRG does not select any diagrammatic channel and hence does not violate such delicate symmetries, which allows us to observe the AFM$^*$ phase.
}
This scenario is equivalent to the simpler Su-Schrieffer-Heeger-Hubbard model (SSH-HM), where a similar transition from CBO to AFM$^*$ phase was found~\cite{Yang2022, Götz2022, Feng2022}.
The transition between the different phases as a function of $\omega$ is depicted along with the important diagrammatic contributions in \cref{fig:omega_sweep}.

\begin{figure*}[!t]
    \centering
    \vspace{-5pt}
    \begin{tikzpicture}[inner sep=0pt, outer sep=0pt]
    \node (a) at (0,0) {
        \includegraphics[width=\linewidth]{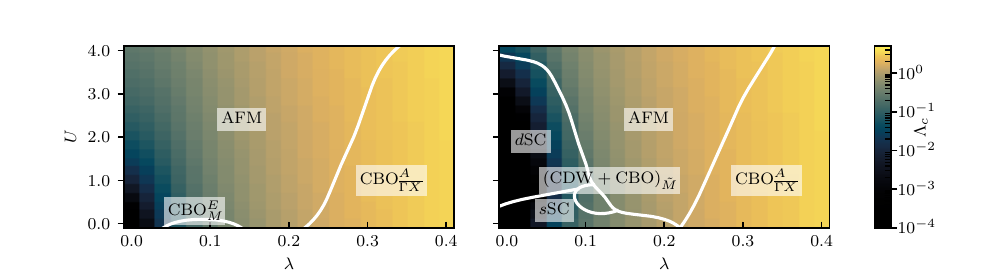}
    };
    \node[anchor=south east, xshift=-90pt, yshift=-15pt] at (a.north) {\small(a), $n = 1.0$};
    \node[anchor=south west, xshift= 55pt, yshift=-15pt] at (a.north) {\small(b), $n = 0.85$};
    \end{tikzpicture}
    \vspace{-15pt}
    \caption{
    Phase diagram in ($U$, $\lambda$) space with $\omega=0.1$, $\alpha^2 = 2/3$ (corresponding to $K_\perp = K_\parallel/2$) at half filling $\langle n \rangle = 1.0$ (a) and slightly away $\langle n \rangle = 0.85$ (b).
    The critical scale $\Lambda_c$ is indicated by color.
    Explanations for the phases are given in the text.
    At half filling and in the case of $U\to 0$, $O(4)$ symmetry is recovered so that the AFM phase gives way to the degenerate AFM$^*$ phase.
    We do not label these $U=0$ phases separately as they occupy an asymptotically small area of the phase diagram.
    }
    \label{fig:U_lambda_phase_diagram}
\end{figure*}

Slightly away from half filling the AFM$^*$ degeneracy between AFM, $s$SC, and CDW is lifted.
Due to the absence of any electronic repulsion, onsite Cooper pairing is the generic FS instability in this regime since perfect FS nesting is absent at finite doping and the particle-hole susceptibility $\chi^H$ is significantly suppressed compared to the $P$-channel vertex.
Consequently, $s$-wave SC dominates the phase diagram in \cref{fig:lambda_alpha_phase_diagram}(d) and leads to a reduction of phase space with CBO and $d$Pom instabilities that are ultimately driven by the geometric series in the $D$-channel.
Notably, the $\tilde M$-type CBO picks up an additional CDW content that is generated by inter-channel fluctuations, so that we term this phase $(\mathrm{CDW}+\mathrm{CBO})_{\tilde M}$.
While this is symmetry allowed already in the $D$-RPA, the limited diagrammatic complexity of the Dyson series is unable to mix different form factor (i.e., CDW and CBO) contributions. Consequently, the shape of the instability is predetermined by the structure of the bare vertex and is bound to nearest-neighbor bond orders.
This limitation is overcome by the FRG leading to a more differentiated microscopic picture of the phase transition, since energetically favorable, symmetry allowed contributions to the order parameter are generated along the flow in a natural way and conventionally reduce the sensitivity with respect to small parameter changes in the bare Hamiltonian~\cite{Profe2024m, Profe2024k}.

\subsubsection{Competition with electronic interactions}

Unlike in most other numerical methods, the inclusion of finite Coulomb repulsion in combination with the electron-phonon coupling as well as doping is straightforward in FRG as outlined in \cref{sec:method}.
We hence add an onsite repulsion $U$ according to \cref{eq:H_e} to the electron-phonon coupling \cref{eq:H_elph} and investigate the resulting ($U$, $\lambda$) phase diagram in \cref{fig:U_lambda_phase_diagram} at fixed $\alpha$ and $\omega$.

In the low $U / \lambda$ regime, the bare static vertex is again dominated by the electron-phonon vertex of \cref{fig:V_eff} with maximum entries on the $\overline{\Gamma X}$ line and we obtain persistent $X$-type CBO orders in this regime in line with the results of \cref{fig:lambda_alpha_phase_diagram}.
However, the picture starts to differ significantly as soon as electronic renormalization effects play a crucial role.
At half filling, the perfect FS nesting at $\bv q = M$ results in a purely electronically driven AFM order that typically persists in the all electronic Hubbard model~\cite{Halboth2000}.
This is well known from the SSH-HM, where any $U>0$ breaks the degeneracy between the AFM, $s$SC, and CDW states in favor of the AFM~\cite{Yang2022, Götz2022, Feng2022}.
The microscopic mechanism can be directly inferred from \cref{fig:omega_sweep}, where the instantaneous electron-electron interaction $U$ acts as the main driver of the AFM instability in the $C$-channel and teams up with the static part of the electron-phonon interaction. The combined interaction supersedes all retarded electron-phonon driven processes, which are dominantly active in the $D$-channels at the small $\omega$ values chosen for \cref{fig:U_lambda_phase_diagram}.
The screening effects of the electron-phonon coupling only influence this picture directly at $U =0$, which is visible in the ($\lambda$, $\alpha$) phase diagram in \cref{fig:lambda_alpha_phase_diagram}(c).

Under doping, the FS nesting and the associated peak of the particle-hole susceptibility at $\bv q = M$ is heavily reduced, such that the AFM state is considerably suppressed.
This sets electronic and phonon mediated screening processes on roughly equal footing and fosters a richer phase diagram.
In the small $\lambda$ regime, where a lowering of the critical scales is achieved by a reduced contribution of the static electron-phonon interaction in the $C$-channel resummation, the AFM fluctuations remain strong but are no longer diverging and promote a d-wave superconducting state ($d\mathrm{SC}$) $\big \langle c^\dagger_{\bv{k} \uparrow} c^\dagger_{-\bv{k} \downarrow} - c^\dagger_{\bv{k} \downarrow} c^\dagger_{-\bv{k} \uparrow} \big \rangle \propto \cos(k_x) - \cos(k_y)$.
At low but finite $U$, the $d$SC is eventually surpassed by conventional $s\mathrm{SC}$, since the onsite Coulomb repulsion is overscreened by attractive electron-phonon coupling at low critical scales.
Notably, we find at finite $U$ a small regime of CDW\,+\,CBO with incommensurate condensation momentum $\bv q = \tilde M$ that was already found in \cref{fig:lambda_alpha_phase_diagram}(d) albeit only at higher $\alpha^2$.
Although the retarded interaction is primarily driven by the $D$-channel series, the CDW\,+\,CBO only emerges through the inclusion of Coulomb repulsion, which suppresses the competing onsite $s$SC.
This highlights the subtle interplay between electron-phonon and electron-electron interactions in the system.

\section{Summary \& Outlook} \label{sec:conclusion}
In summary, we demonstrate how to unify the electron-phonon coupled picture of Peierls transitions with the purely electronic viewpoint in arbitrary minimal lattice models with acoustic phonons.
In the electronic picture, we detail the use of the FRG to include vertex corrections as well as pair fluctuations on equal footing.
This allows us to describe a multitude of electron-phonon coupling driven phases beyond Peierls' instability.
Furthermore, the reformulation in the electronic sector enables the study of competition between Coulomb repulsion and electron-phonon coupling, resulting in various intricate phase diagrams.
Extracting the leading diagrammatic content of the respective phases stresses the importance of vertex corrections and pair fluctuations in significant regions of phase space, \emph{a posteriori} justifying FRG as inevitable tool when characterizing electron-phonon coupled models with and even without electron-electron repulsion.
Despite its subleading role for Fermi surface instabilities~\cite{salmhofer2001fermionic}, the renormalization of the electronic band structure through electron-phonon coupling will also be considered in the future in order to increase the quantitative nature of the present framework.

For the Hubbard-model with acoustic phonons on a square lattice, the phononic picture uncovers a competition between two classes of charge bond order: $X$- or $M$-type.
By the FRG treatment we find that a degenerate AFM/$s$SC/CDW state known from the Su-Schrieffer-Heeger-Hubbard-model dominates large parts of the phase diagram, especially for weak electron-phonon coupling.
Upon introducing a Hubbard repulsion, the diagrammatic content from electron-phonon coupling intertwines with the electronic interactions and gives rise to a complex phase diagram of competition between $d$-wave/$s$-wave SC, AFM, and the aforementioned two types of CBO.

The methodological advancement in conjunction with the quality of the approximations~\cite{Al-Eryani2025} of this work will be of particular merit in the broad context of unbiased \emph{ab initio} characterization of intertwined charge, spin, and pairing instabilities.
In the future, we plan to extend the \textsc{divERGe} library~\cite{10.21468/SciPostPhysCodeb.26,10.21468/SciPostPhysCodeb.26-r0.5} by a generic interface to electron-phonon coupled first principle models, similar to, e.g., the \textsc{elphmod} package~\cite{berges2025elphmod} but for electronic FRG.
Notably, the generality of the approach presented here enables us to treat arbitrary electron-phonon coupled models, including, e.g., spin-orbit coupling, optical \emph{and} acoustic phonon modes, multi-orbital and multi-site setups, topological systems, etc.
We thereby aim to bridge the gap between \emph{ab initio} downfolding and FRG approaches to correlated electron-phonon systems, allowing for prompt application to various highly relevant material platforms.
Examples include, among others,
(i)~the mechanism for spin- and charge order in the Kagome metal CsCr\textsubscript{3}Sb\textsubscript{5}, which is known to be structurally unstable in DFPT \emph{and} strongly correlated in the electronic sector~\cite{liu2024superconductivity, xu2025altermagnetic},
(ii)~the interplay of (phononic) CDW orders and unconventional electronic correlations such as loop current order or superconductivity in Vanadium based (AV\textsubscript3Sb\textsubscript5) Kagome metals~\cite{guo2024correlated, enzner2025phonon, zhan2025loop},
(iii)~the origin of superconductivity in rhombohedral and twisted graphene multilayers~\cite{cao2018unconventional, zhou2022isospin, chou2022acousticphononmediated},
(iv)~peculiarities about surface superconductivity in topological insulators~\cite{maeland2025mechanism, maeland2025phonon, klebl2026surface},
(v)~charge- and spin order in Nickelate superconductors and the interplay with superconductivity~\cite{Zhang2020, Li2025, Khasanov2025}, 
(vi)~the competition of charge and excitonic order in TiSe\textsubscript2~\cite{Cercellier2007, Pashov2025} as well as chiral charge order and chiral phonons in TiTe\textsubscript2~\cite{Ren2023, Zhang2024}, (vii)~pairing at the LAO/STO interface~\cite{gariglio2015electron, scheurer2015topological},
and (viii)~unconventional CDW materials in general~\cite{Johannes2008, diego2026fermi}.

\begin{acknowledgments}
We thank Aiman~Al-Eryani, Michael~Scherer, Xianxin~Wu, and Jun~Zhan for stimulating discussions.
Further, we acknowledge correspondence with Stefan~Enzner and Kristian~Mæland.
This work is funded by the Deutsche Forschungsgemeinschaft (DFG, German Research Foundation) through the Würzburg-Dresden Cluster of Excellence \emph{ctd.qmat} – Complexity, Topology and Dynamics in Quantum Matter (EXC 2147, project-id 390858490).
We acknowledge HPC resources provided by the Erlangen National High Performance Computing Center (NHR@FAU) of the Friedrich-Alexander-Universität Erlangen-Nürnberg (FAU).
NHR funding is provided by federal and Bavarian state authorities.
NHR@FAU hardware is partially funded by the DFG 440719683.
MD acknowledges support from the Studienstiftung des deutschen Volkes.
\end{acknowledgments}

\bibliography{bibliography}
\end{document}